\documentclass[aps,prl,nobibnotes,twocolumn,superscriptaddress,bibliography]{revtex4-2}

\usepackage{color}
\usepackage{amsmath}
\usepackage{amssymb}
\usepackage{graphicx}
\usepackage{amsfonts}
\usepackage{mathrsfs}
\usepackage{bm}
\usepackage{xspace}
\usepackage{epstopdf}
\usepackage{dcolumn}
\usepackage{longtable}
\usepackage{multirow}
\usepackage{float}
\usepackage{comment}
\usepackage{lineno}

\usepackage[colorlinks=true, letterpaper=true, pdfstartview=FitV,  linkcolor=blue, citecolor=blue, urlcolor=blue]{hyperref}

\makeatother
\begin{document}
\title{Bulk-Fermi-Arc Transition Induced Large Photogalvanic Effect in Weyl Semimetals}
\author{Jin Cao}
\altaffiliation{These authors contributed equally to this work.}
\affiliation{Centre for Quantum Physics, Key Laboratory of Advanced Optoelectronic Quantum Architecture and Measurement (MOE), School of Physics, Beijing Institute of Technology, Beijing, 100081, China}
\affiliation{Beijing Key Lab of Nanophotonics \& Ultrafine Optoelectronic Systems, School of Physics, Beijing Institute of Technology, Beijing, 100081, China}
\author{Maoyuan Wang}
\altaffiliation{These authors contributed equally to this work.}
\affiliation{Centre for Quantum Physics, Key Laboratory of Advanced Optoelectronic Quantum Architecture and Measurement (MOE), School of Physics, Beijing Institute of Technology, Beijing, 100081, China}
\affiliation{Beijing Key Lab of Nanophotonics \& Ultrafine Optoelectronic Systems, School of Physics, Beijing Institute of Technology, Beijing, 100081, China}
\author{Zhi-Ming Yu}
\email{zhiming\_yu@bit.edu.cn}
\affiliation{Centre for Quantum Physics, Key Laboratory of Advanced Optoelectronic Quantum Architecture and Measurement (MOE), School of Physics, Beijing Institute of Technology, Beijing, 100081, China}
\affiliation{Beijing Key Lab of Nanophotonics \& Ultrafine Optoelectronic Systems, School of Physics, Beijing Institute of Technology, Beijing, 100081, China}
\author{Yugui Yao}
\email{ygyao@bit.edu.cn}
\affiliation{Centre for Quantum Physics, Key Laboratory of Advanced Optoelectronic Quantum Architecture and Measurement (MOE), School of Physics, Beijing Institute of Technology, Beijing, 100081, China}
\affiliation{Beijing Key Lab of Nanophotonics \& Ultrafine Optoelectronic Systems, School of Physics, Beijing Institute of Technology, Beijing, 100081, China}

\begin{abstract}
The surface Fermi arc, as a hallmark of Weyl semimetals (WSMs), has
been well known in current research, but it remains a challenge to
unveil novel phenomena associated with the Fermi arc. Here, we predict
a heretofore unrecognized process in WSMs, namely, the photoinduced
transition between the bulk states and the Fermi arc. We find this
process is significant and can lead to a large effective three-dimensional
shift current on the boundaries with the Fermi arc in wide terahertz
range. Moreover, due to the low symmetry of the boundaries, the surface
photogalvanic effect predicted here can appear in a large class of
WSMs that do not have bulk shift current. Hence, our work not only
unveils a hidden photogalvanic effect in WSMs but also suggests that
all the WSMs are promising material candidates for developing efficient
terahertz photodetectors.
\end{abstract}

\maketitle

\paragraph{\textcolor{blue}{Introduction.}}

The photogalvanic effect refers to the generation of dc electric current
in a material illuminated by light~\citep{boyd2020nonlinear_start}.
It has been attracting intensive interest in condensed matter physics,
due to the promising application on photodetectors and solar cells
beyond pn junction structure~\citep{
    prl2012_Young_dft_1_bpve,prl2012_Young_dft_2_bpve,prl2014_Somma_bpve,
    nc2017_Nakamura_exp_bpve,prb2017_Wang_sc_dft_bpve,prl2018_GaAs_mat_bpve,
    SciAdv2019_Burgereaau_exp_bpve,prb2019_Fregoso_dft_injection_bpve,SciAdv2019_Wang_bpve}.
In addition to technological application, the photogalvanic effect
also provides a basic mechanism to probe various geometric quantities
of systems, such as the Berry curvature, quantum metric and Christoffel
symbols~\citep{
    prb2000_sipe_formula_bpve,prb2016_JMoore_formula_bpve,
    SciAdv2016_nagaosa_optgeometry,Advmat2017_nagaosa_optgeometry,
    prb2017_Benjamin_weylopt,prl2018_sipe_jerkcurrent_bpve,
    prb2019_Jmoore_diagram_formula_bpve,prx2020_nagaosa_formula_bpve,
    prr2020_Yan_optgeometry}.

Generally, the dominant dc response of materials under a monochromatic
light characterized by
$\boldsymbol{A}\left(t\right)=\boldsymbol{A}\left(\omega\right)e^{-i\omega t}+c.c.$
is quadratic, and the shift current, a representative photogalvanic
effect, can be expressed as~\citep{prb2000_sipe_formula_bpve}
\begin{eqnarray}
j^{a} & = & \sigma_{bc}^{a}\left(\omega\right)E^{b}\left(\omega\right)E^{c}\left(-\omega\right),\label{eq:current}
\end{eqnarray}
with $\sigma_{bc}^{a}$ the third-rank shift conductivity tensor and
$\boldsymbol{E}\left(\omega\right)=i\omega\boldsymbol{A}\left(\omega\right)$.
Here, we have adopted the Einstein summation convention and the roman
letters $a,b,\cdots$ denote Cartesian indexes. Clearly, the tensor
$\sigma_{bc}^{a}$ is constrained by the (magnetic) point group of
the systems. For example, when the system has spatial inversion symmetry
$\mathcal{P}$, $\sigma_{bc}^{a}$ should vanish, as both $\boldsymbol{j}$
and $\boldsymbol{E}$ change sign under $\mathcal{P}$. Hence, the
studies of shift current effect generally focus on noncentrosymmetric
materials~\citep{nm2020_Liu_review_weylopt,review2020_nagaosa}.

Many efforts have been devoted to searching material candidates with
large shift current effect. For example, A. M. Cook \emph{et al.}~
\citep{nc2017_Cook_Design_principles_bpve} predicted that the shift
current in the semiconductors with semi-Dirac type of Hamiltonian
is large and may compete with conventional solar cells based on pn
junction. T. Rangel \emph{et al.} \citep{prl2017_rangel_GeSn_monochalcogenides_bpve}
confirmed it by performing first-principles calculations on single-layer
monochalcogenides with such low-energy Hamiltonian and found the effective
three-dimensional shift conductivity in these materials is larger
than that in many other polar systems.

Currently, the exploration is extended to topological WSMs \citep{
    prl2016_Nagaosa_weylopt,prb2016_Taguchi_weylopt,
    nc2017_Juan_exp_QCPGE_optgeometry,prb2017_lee_tiltedWeyl_weylopt,
    prb2017_Levchenko_weylopt,np2017_Ma_exp_weylopt,arxiv2017_Yang_weylopt,
    np2017_Wu_exp_SHG_weylopt,prb2018_Flicker_multifold_quantizedCPGE_optgeometry,
    prb2018_Golub_weylopt,prl2018_Nagaosa_weylopt,prb2018_Yan_TaAs_weylopt,
    nm2019_Xu_TaAs_weylopt,nm2019_Ma_exp_typeII_optgeometry,
    SciAdv2020_exp_QCPGE_optgeometry,arxiv2021_Jennifer_multifold_weylopt}.
For WSMs, the conduction band and the valence band form Weyl points
in the bulk, around which all the geometric quantities become divergent,
and the topological surface Fermi arc state appears on the boundaries
\citep{prb2011_wan_YIrO_weyl,rmp2018_Ashvin_Weyl}. Since the shift
current is closely related to the shift vector and the Christoffel
symbols \citep{prb2000_sipe_formula_bpve,prx2020_nagaosa_formula_bpve},
one can expect that the shift current in WSM may be significant. Moreover,
the gapless feature makes WSMs an ideal choice for designing terahertz
photodetectors \citep{prb2017_lee_tiltedWeyl_weylopt,review2020_nagaosa}.
However, as aforementioned the shift current is constrained by the
symmetries of system. Besides $\mathcal{P}$ symmetry, an emergent
SO(3) rotational symmetry, which typically exists for the Weyl points
at high-symmetry points under certain space group
symmetries \citep{arxiv2021_yu_encyclopedia,nm2018_chang_chiral_crystals},
would also cause the shift current to vanish \citep{prx2020_nagaosa_formula_bpve}.
Thus, a large class of WSMs are forbidden by symmetries to have net
bulk shift current. There also exist few works that studied the photogalvanic
effect on the boundary of the topological materials, focusing on the
photocurrent solely induced by the topological surface states \citep{
    prl2020_cpge_RhSi_surfres,prb2011_Hosur_cpgeTI_surfres,
    nc2016_surface_Bi2Se3_surfres,prb2017_Kim_puresurface_surfres,
    nc2019_Wang_exp_surfres,am2020_Chi_nodalline_surfres}.
But the shift current effect there generally are not significant.

\begin{figure}
\begin{centering}
\includegraphics[width=8.5cm]{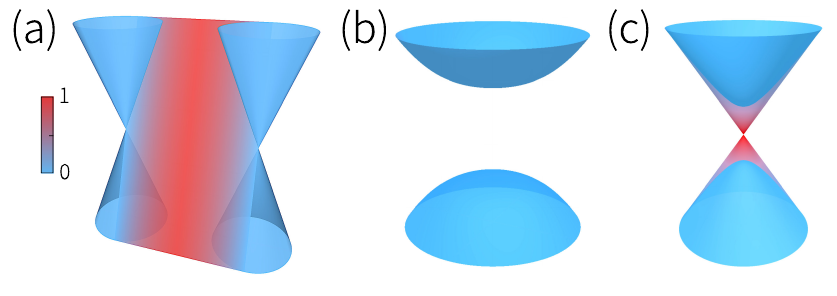}
\par\end{centering}
\caption{\label{Fig.1} Three typical band structures. (a) Weyl semimetal with
surface Fermi arc. (b) Trivial insulator and (c) topological insulator
with topological surface state. The blue surfaces denote bulk bands.
The surface states in (a) and (c) are attached to the bulk bands.
The color map in surface states indicates the weight of projection
onto surface. Considerable  transition between bulk and surface states would
widely appears in (a), while can not happen for (b) and happen
in (c) for certain doping.}
\end{figure}

In this work, we show that large surface shift current can be generated
by a unique and heretofore unrecognized process in WSMs, namely, the
transition between bulk states and the surface Fermi arc. This process
is termed as bulk-Fermi-arc transition. A key observation is that
the surface Fermi arc is always attached to the bulk Weyl cones at
any energy \citep{prb2011_wan_YIrO_weyl}, and there exists considerable
overlap between the bulk states and the part of the Fermi arc attached
to the bulk \citep{prb2015_pan_valley}, as illustrated in Fig. \ref{Fig.1}(a).
Hence, considerable bulk-Fermi-arc transition should widely occur in WSMs.

We first establish a local second-order response formula, and then
use it to study the surface shift current of a WSM with $\mathcal{P}$
symmetry and only one pair of Weyl points. In this WSM model, both
the bulk shift current and the surface shift current solely from the
surface Fermi arc are zero. Surprisingly, we find that significant
surface shift current effect can occur and only occur on the boundaries
with Fermi arc, indicating that this considerable large shift current
is induced by the bulk-Fermi-arc transition. Particularly, the absolute
value of the effective three-dimensional shift current induced by
the bulk-Fermi-arc transition can exceed $100\ \mu$A/V$^{2}$ in
a wide terahertz range, due to the gapless spectrum between the bulk
and the Fermi arc {[}see Fig.~\ref{Fig.1}(a){]}. Such large shift
current is comparable with that in the single-layer monochalcogenides
\citep{prl2017_rangel_GeSn_monochalcogenides_bpve} and the bulk shift
current in the previously studied topological semimetals \citep{prb2018_Yan_TaAs_weylopt}.
Moreover, due to the low symmetry on the boundary, the surface shift
current effect predicted here generally would appear in all kinds
of the WSMs, including a large class of WSMs that do not have net
shift current in bulk. This greatly relaxes the applied restrictions
of WSM in terahertz photodetectors.

\paragraph{\textcolor{blue}{Local photocurrent response.}}

We first establish a general formula to describe the local photocurrent
response to study spatially resolved shift current effect. The space
dependent current operator generally can be written as \citep{mahan1990many}
\begin{eqnarray}
J_{a}\left(\boldsymbol{r}\right) & = & -\frac{e}{2V}\sum_{\mu\nu,\boldsymbol{r}^{\prime}}v_{\mu\boldsymbol{r},\nu\boldsymbol{r}^{\prime}}^{a}d_{\mu\boldsymbol{r}}^{\dagger}d_{\nu\boldsymbol{r}^{\prime}}+h.c.,
\end{eqnarray}
which counts all of the currents flowing from $\boldsymbol{r}^{\prime}$
to $\boldsymbol{r}$. Here, $V$ is the volume of system, $e$ denotes
the charge carried by an electron, $d_{\mu\boldsymbol{r}}^{\dagger}$
($d_{\mu\boldsymbol{r}}$) creates (annihilates) a localized Wannier
state $w_{\mu\boldsymbol{r}}$ at $\boldsymbol{r}$ with basis orbit
$\mu$, and
$v_{\mu\boldsymbol{r},\nu\boldsymbol{r}^{\prime}}^{a}=\frac{1}{2m}\left\langle w_{\mu\boldsymbol{r}}|-i\partial_{a}|w_{\nu\boldsymbol{r}^{\prime}}\right\rangle $
with $m$ the mass of the electrons. The eigenstates are constructed
from the Wannier states by $c_{n}^{\dagger}=\sum_{\mu\boldsymbol{r}}U_{\mu\boldsymbol{r},n}d_{\mu\boldsymbol{r}}^{\dagger}$
with $n$ a combined index denoting energy band and momentum (if the
system has translation symmetry), it then follows
\begin{eqnarray}
J_{a}\left(\boldsymbol{r}\right) & = & -\frac{e}{V}\sum_{mn}v_{mn}^{a}\left(\boldsymbol{r}\right)c_{m}^{\dagger}c_{n},
\end{eqnarray}
where $v_{mn}^{a}\left(\boldsymbol{r}\right)$ is the local velocity
operator matrix, defined as
\begin{eqnarray}
v_{mn}^{a}\left(\boldsymbol{r}\right) & \equiv & \frac{1}{2}\sum_{\mu\nu,\boldsymbol{r}^{\prime}}U_{m,\mu\boldsymbol{r}}^{\dagger}v_{\mu\boldsymbol{r},\nu\boldsymbol{r}^{\prime}}^{a}U_{\nu\boldsymbol{r}^{\prime},n}+h.c..\label{eq:v_local}
\end{eqnarray}

Consider a slab model with $y$-direction being confined and assuming
the model is uniformly illuminated by monochromatic light of frequency
$\omega$. The applied light (electric) field can be introduced into
the Hamiltonian by velocity-gauge approach, $H^{\prime}=-\int d\boldsymbol{r}\,\boldsymbol{J}\left(\boldsymbol{r}\right)\cdot\boldsymbol{A}\left(\boldsymbol{r},t\right)$.
According to the standard perturbation procedure, the local shift
conductivity can be established as (see supplemental materials (SM)~\citep{Supplemental_Materials})
\begin{eqnarray}
 &  & \sigma_{bc}^{a}\left(\omega;y\right)=-\frac{\pi e^{3}}{2L\omega^{2}}\mathrm{Im}\int\frac{dk_{x}dk_{z}}{\left(2\pi\right)^{2}}\sum_{y^{\prime},y^{\prime\prime}}\sum_{lmn;\pm\omega}\frac{f_{nl}}{\varepsilon_{lm}}\nonumber \\
 &  & \quad\times v_{lm,y}^{a}\left(v_{mn,y^{\prime}}^{b}v_{nl,y^{\prime\prime}}^{c}+v_{mn,y^{\prime\prime}}^{c}v_{nl,y^{\prime}}^{b}\right)\delta\left(\varepsilon_{ln}-\hbar\omega\right),\label{eq:lpge_abc_yyy}
\end{eqnarray}
where $L$ is the thickness of the slab, $f_{nl}=f_{n}-f_{l}$ and
$\varepsilon_{lm}=\varepsilon_{l}-\varepsilon_{m}$ are the occupation
and the energy differences between the two states involved in the
optical transition, $f_{n}$ is the Fermi-Dirac distribution. Since
Eq. (\ref{eq:lpge_abc_yyy}) has the form of Fermi golden rule, the
shift current effect is a interband effect.

For top surface, the effective three-dimensional shift conductivity
may be defined as
\begin{eqnarray}
\Sigma_{bc}^{a}\left(\omega\right) & = & \frac{1}{l}\int_{L/2-l}^{L/2}dy\:\sigma_{bc}^{a}\left(\omega;y\right),\label{eq:surcon}
\end{eqnarray}
where $l\ll L$ is the distance measured from the top surface at $L/2$.
The conductivity for bottom surface can be similarly defined. When
the system has $\mathcal{P}$ symmetry, the surface shift conductivity
for the top and bottom surfaces would take opposite values. While
the value of $\Sigma_{bc}^{a}\left(\omega\right)$ has a dependence
on $l$, the qualitative behaviors of $\Sigma_{bc}^{a}\left(\omega\right)$
would be robust against the choice of $l$.

\paragraph{\textcolor{blue}{Weyl model.}}

Since our goal is to demonstrate the existence of the surface photocurrent
induced by bulk-Fermi-arc transition, we take a simplest WSM model
with only two conventional Weyl points without energy tilt at Fermi
level. The essential physics learned here applies to general WSMs
and other topological semimetals with surface Fermi arc. We also assume
the system has $\mathcal{P}$ symmetry to exclude the bulk photocurrent.
This indicates the WSM model has to break time-reversal symmetry ($\mathcal{T}$),
while a combined operator $\mathcal{OT}$ with $\mathcal{O}$ a certain
spatial operator may be persevered. Since there exists (at most) only
one surface Fermi arc on each boundary of system, the surface shift
current solely from the Fermi arc would be zero.

Consider a tight-binding model defined on a cubic lattice
\begin{eqnarray}
\mathcal{H}_{\mathrm{W}}\left(\boldsymbol{k}\right) & = & \left[\Delta+t_{1}\left(\cos k_{x}+\cos k_{y}\right)+t_{2}\cos k_{z}\right]\sigma_{z}\nonumber \\
 &  & +t_{3}\left(\sin k_{x}\sigma_{x}+\sin k_{y}\sigma_{y}\right),\label{eq:weylhamTB}
\end{eqnarray}
where $\sigma_{i}$'s are the Pauli matrixes, $t_{1}=t_{2}=1.0\,\mathrm{eV}$,
$t_{3}=0.25\,\mathrm{eV}$ and $\Delta=\Delta_{0}-2t_{1}-t_{2}$ with
$\Delta_{0}=0.5\,\mathrm{eV}$. This lattice model (\ref{eq:weylhamTB})
has only two Weyl points on the $k_{z}$ axis at $k_{z}^{\pm}=\pm\arccos(2\Delta_{0}/t_{1})$.
These two Weyl points are conventional linear Weyl points with Chern
number $C=\pm1$. In the bulk, the symmetries of the lattice model
are generated by $\mathcal{P}$ symmetry, a mirror $\mathcal{M}_{z}$,
a fourfold rotation $\mathcal{C}_{4z}$ and a combined operator $\mathcal{C}_{2y}\mathcal{T}$,
which respectively can be represented as $\mathcal{P}=\sigma_{z}$,
$\mathcal{M}_{z}=i\sigma_{0}$, $\mathcal{C}_{4z}=\frac{\sigma_{0}+i\sigma_{3}}{\sqrt{2}}$
and $\mathcal{C}_{2y}\mathcal{T}=\mathcal{K}$.

\begin{figure}
\begin{centering}
\includegraphics[width=8.5cm]{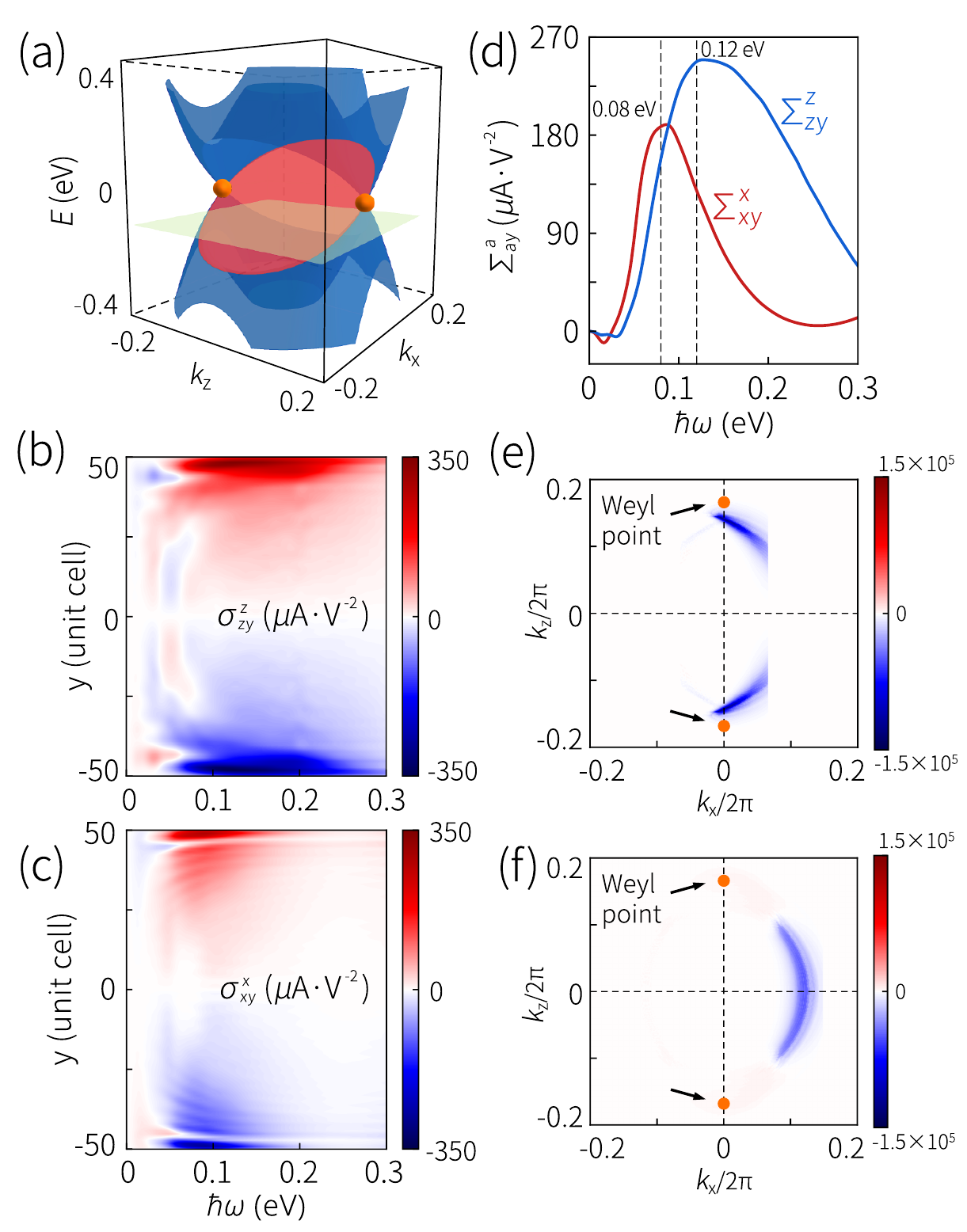}
\par\end{centering}
\caption{\label{Fig.2} (a) Band structure of the model (\ref{eq:weylhamTB})
at $k_{y}=0$ plane (blue surfaces), together with the surface Fermi
arc states at (010) surface (red surface). Green surface denotes the
Fermi level. (b,c) show the local shift conductivity $\sigma_{zy}^{z}\left(\omega;y\right)$
and $\sigma_{xy}^{x}\left(\omega;y\right)$ for the (010) slab, where
significant surface enhancement can be observed. We set $E_{F}=-0.1\,\mathrm{eV}$
in (b), $E_{F}=-0.2\,\mathrm{eV}$ in (c). (d) shows the two surface
shift conductivities which are obtained from (b) and (c) with $l=4\,L_{s}$.
(e-f) The distribution of $\sigma_{zy}^{z}\left(\omega;y=L/2\right)$
and $\sigma_{xy}^{x}\left(\omega;y=L/2\right)$ in the (010) surface
BZ. We set $E_{F}=-0.1\,\mathrm{eV}$ and $\hbar\omega=0.12\,\mathrm{eV}$
in (e), and $E_{F}=-0.2\,\mathrm{eV}$ and $\hbar\omega=0.08\,\mathrm{eV}$
in (f).}
\end{figure}

The calculated band structure from Eq. (\ref{eq:weylhamTB}) is plotted
in Fig. \ref{Fig.2}(a), together with the surface Fermi arc on the
(010) surface. One observes that the two Weyl points locate at $k_{z}^{\pm}\boldsymbol{\hat{z}}$
points respectively. Moreover, the bands of the two Weyl points are
connected at a higher (lower) energy to form a saddle surface around
$k_{x}$ axis. At low energy, the surface Fermi arc is attached to
the Weyl cones, as discussed above {[}see Fig.~\ref{Fig.1}(a){]}.
Besides, the Fermi arc would also merge into the bulk saddle surface
at higher (lower) energy. Thus, there generally exist two different
photoinduced bulk-Fermi-arc transitions in WSMs, namely, one is the
transition from the Weyl points to the Fermi arc and other is from
the saddle surface to the Fermi arc.

\paragraph{\textcolor{blue}{Surface shift current.}}

We then study the shift current on the (010) surface of the lattice
model (\ref{eq:weylhamTB}) under an uniform irradiation of a linearly
polarized light. The nonzero components of the surface shift conductivity
tensors are determined by the magnetic point group of the (010) surface,
which are generated by $\mathcal{M}_{z}$ and $\mathcal{C}_{2y}\mathcal{T}$.
Then the shift conductivity tensors, that are odd under the mirror
$\mathcal{M}_{z}$, namely, $\sigma_{xz}^{x}$, $\sigma_{yz}^{x}$,
$\sigma_{xx}^{z}$, $\sigma_{yy}^{z}$, $\sigma_{zz}^{z}$, $\sigma_{xy}^{z}$,
$\sigma_{xz}^{z}$, and $\sigma_{yz}^{z}$ vanish. Similarly, $\sigma_{xx}^{x}$,
$\sigma_{yy}^{x}$, $\sigma_{zz}^{x}$, $\sigma_{xz}^{z}$ are excluded
by $\mathcal{C}_{2y}\mathcal{T}$. Thus, there are only four symmetry
allowed tensors and two of them are independent, which are $\sigma_{xy}^{x}=\sigma_{yx}^{x}$
and $\sigma_{zy}^{z}=\sigma_{yz}^{z}$.

Using the established formula (\ref{eq:lpge_abc_yyy}), we calculate
the local shift conductivities $\sigma_{xy}^{x}\left(\omega;y\right)$
and $\sigma_{zy}^{z}\left(\omega;y\right)$ for a (010) slab with
a thickness of $L=101\,L_{s}$. Here, $L_{s}$ is the thickness of
a unit cell. The obtained results of the local shift conductivities
regarding with $\omega$ are shown in Fig.~\ref{Fig.2}(b) and \ref{Fig.2}(c).
One finds that both $\sigma_{xy}^{x}\left(\omega;y\right)$ and $\sigma_{zy}^{z}\left(\omega;y\right)$
are finite for a generic layer of the slab, as a generic layer has
the same symmetry conditions as the (010) surface. Remarkably, the
two conductivities not only are sizable but also can feature strong
surface enhancement behavior for certain frequencies, leading to significant
photogalvanic effect on the (010) surface. For clarity, we calculate
the top surface shift conductivities $\Sigma_{xy}^{x}\left(\omega\right)$
and $\Sigma_{zy}^{z}\left(\omega\right)$ with $l=4\,L_{s}$, and
find they can be larger than $100\,\mu\mathrm{A}/\mathrm{V}^{2}$
in a wide frequency range, as shown in Fig.~\ref{Fig.2}(d). Such
large surface shift conductivities to our best knowledge has never
been reported before. Moreover, the peak of both $\Sigma_{xy}^{x}\left(\omega\right)$
and $\Sigma_{zy}^{z}\left(\omega\right)$ appear around terahertz
range ($\hbar\omega\sim100\,\mathrm{meV}$), indicating that the surface
photogalvanic effect predicted here can be used to design efficient
infrared and terahertz photodetectors.

To further study the significant surface photogalvanic effect, we
calculate the distribution of the local conductivities $\sigma_{zy}^{z}\left(\omega;y=L/2\right)$
and $\sigma_{xy}^{x}\left(\omega;y=L/2\right)$ that exhibit strong
surface enhancement behavior in the 2D Brillouin zone (BZ) of the
(010) slab. The results are shown in Fig.~\ref{Fig.2}(e) and \ref{Fig.2}(f).
Interestingly, the distribution for both conductivities concentrates around the Fermi arc, indicating the enhanced surface shift current
is closely related to the Bulk-Fermi-arc transition. However, the large contribution for $\sigma_{zy}^{z}$ comes from the transition
between the two Weyl points and the Fermi arc {[}see Fig.~\ref{Fig.2}(e){]},
while for $\sigma_{xy}^{x}$ is from the transition between the saddle
surface and the arc {[}see Fig.~\ref{Fig.2}(f){]}. This difference
will lead to completely different low-energy behavior of the two surface
conductivities.

\begin{figure}
\begin{centering}
\includegraphics[width=8.6cm]{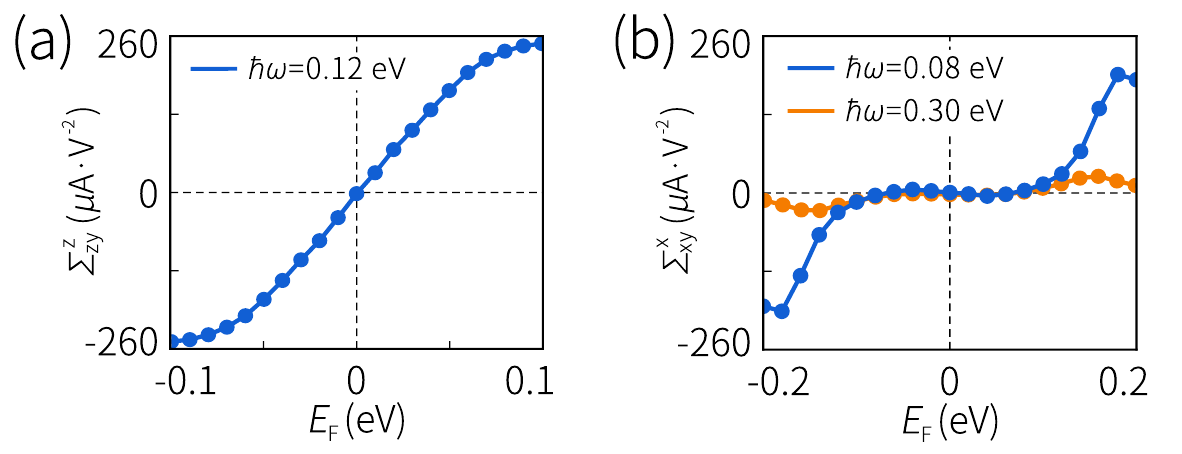}
\par\end{centering}
\caption{\label{Fig.3} The (010) surface shift conductivities (a) $\Sigma_{zy}^{z}$ and (b) $\Sigma_{xy}^{x}$ of the Weyl model (\ref{eq:weylhamTB}) versus the Fermi level. }
\end{figure}

When the Fermi level exactly locates at Weyl points $E_{\mathrm{F}}=0$,
the model (\ref{eq:weylhamTB}) has an emergent particle-hole symmetry
$\mathcal{C}=\sigma_{x}\mathcal{K}$ with
$\mathcal{C}\mathcal{H}_{\mathrm{W}}\left(\boldsymbol{k}\right)\mathcal{C}^{-1}=-\mathcal{H}_{\mathrm{W}}\left(-\boldsymbol{k}\right)$,
in such case all the shift conductivities would vanish, as the particle-hole
symmetry reverses the direction of the photocurrent \citep{prx2020_nagaosa_formula_bpve}.
Deviating from the neutral filling point, the absolute value of $\Sigma_{zy}^{z}\left(\omega\right)$
will increase rapidly {[}see Fig.~\ref{Fig.3}(a){]}. In contrast,
$\Sigma_{xy}^{x}\left(\omega\right)$ will still be almost vanishing
at low energy $\left|E_{\mathrm{F}}\right|<0.1\,\mathrm{eV}$ for
any frequency {[}see Fig.~\ref{Fig.3}(b){]}. The vanishing $\Sigma_{xy}^{x}\left(\omega\right)$
is because that at low energy, the Fermi arc near $k_{x}$ axis is
mainly localized on the surface and has vanishing overlap with the
bulk state {[}see Fig.~\ref{Fig.2}(a){]}. $\Sigma_{xy}^{x}\left(\omega\right)$
becomes significant only when the Fermi level approaches the bulk
saddle surface and the frequency $\hbar\omega$ is at terahertz range,
as shown in Fig. \ref{Fig.3}(b). This again demonstrates that only
the part of the Fermi arc connected to the bulk states can have a
considerable overlap with the bulk states, and then the bulk-Fermi-arc
transition is promoted.

\begin{figure}[b]
    \begin{centering}
    \includegraphics[width=8.5cm]{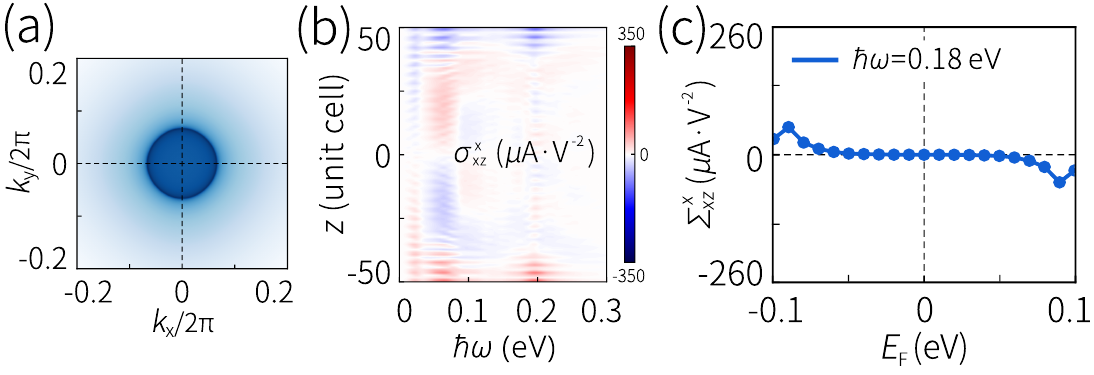}
    \par\end{centering}
    \caption{\label{Fig.4} (a) Surface spectra on (001) surface of the lattice model (\ref{eq:weylhamTB}).
    (b) The local shift conductivity $\sigma_{xz}^{x}\left(\omega;z\right)$
    for the (001) slab. In (a) and (b), we set $E_{\mathrm{F}}=-0.1\,\mathrm{eV}$. (c)   The (001) surface shift conductivity versus the Fermi level.}
\end{figure}

For comparison, we also calculate the local shift conductivity for
the (001) slab of the WSM model (\ref{eq:weylhamTB}) under same parameters.
For (001) surface, only one shift current conductivity $\sigma_{xz}^{x}\left(\omega;z\right)$
is independent (see SM~\citep{Supplemental_Materials}). Unlike the (010) surface, the (001) surface does not have a Fermi arc because the two Weyl points are projected to the same point in the BZ of the (001) surface [see Fig.~\ref{Fig.4}(a)]. We find that for different $\hbar\omega$ and $E_{\mathrm{F}}$, the $\sigma_{xz}^{x}\left(\omega;z\right)$
does no feature significant surface enhancement behavior, and the
obtained surface shift conductivity is much smaller than that in (010)
surface, as shown in Fig.~\ref{Fig.4}(b) and \ref{Fig.4}(c). These
results unambiguously demonstrates the significant surface photogalvanic
effect in WSMs is induced by the bulk-Fermi-arc transition.

\begin{figure}
\begin{centering}
\includegraphics[width=8.5cm]{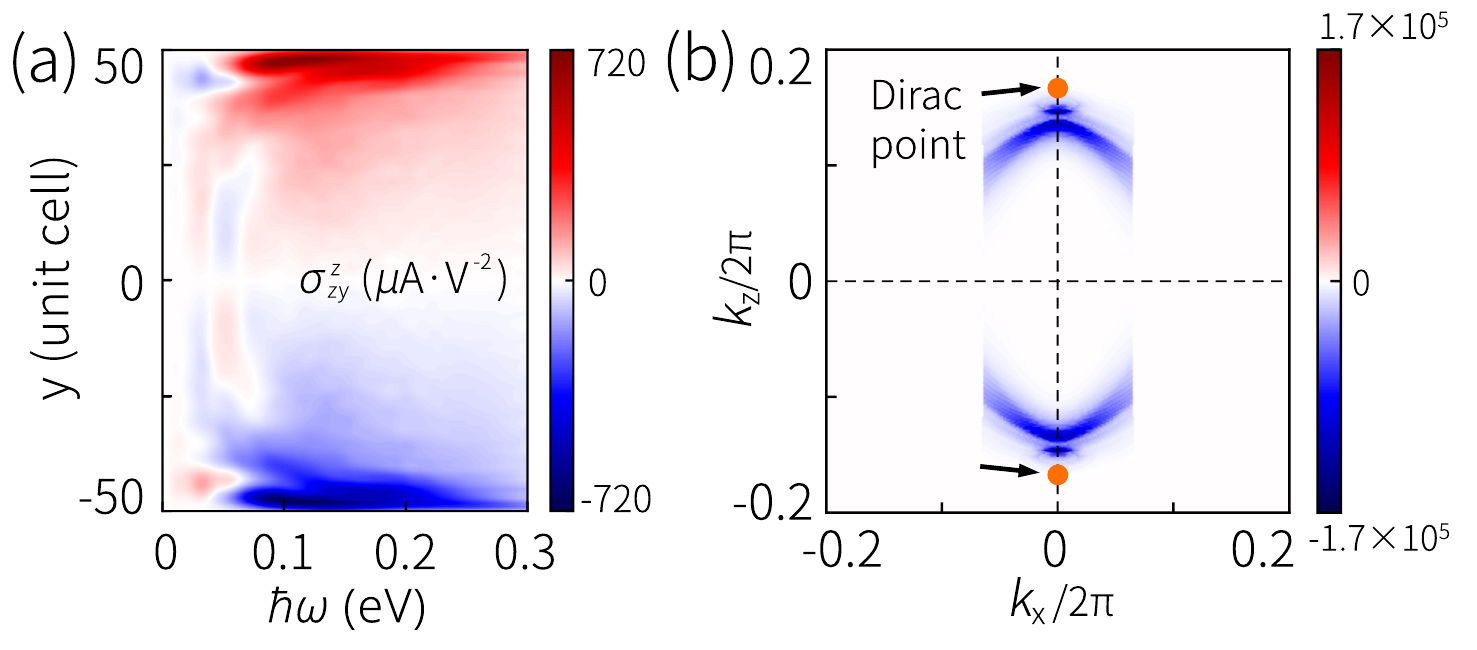}
\par\end{centering}
\caption{\label{Fig.5} (a) The local shift conductivity $\sigma_{zy}^{z}\left(\omega;y\right)$
for the (010) slab of the Dirac model (\ref{eq:DirachamTB}). (b)
shows the distribution of $\sigma_{zy}^{z}\left(\omega;y=L/2\right)$
in the (010) surface BZ. We set $c_{1}=c_{2}=0.1\,\mathrm{eV}$, $E_{\mathrm{F}}=-0.1\,\mathrm{eV}$
in (a) and (b), and set $\hbar\omega=0.12\,\mathrm{eV}$ in (b).}
\end{figure}

\paragraph{\textcolor{blue}{Discussions.}}

In this work, we use a WSM model to demonstrate the existence of the
large surface photovoltaic effect induced by the bulk-Fermi-arc transition.
Actually, this effect would generally exist in all the topological
semimetals with surface Fermi arc. As an example, we consider a Dirac
semimetals with $\mathcal{P}$ and $\mathcal{T}$,
\begin{eqnarray}
\mathcal{H}_{\mathrm{D}}\left(\boldsymbol{k}\right) & = & \begin{bmatrix}\mathcal{H}_{\mathrm{W}}\left(\boldsymbol{k}\right) & h_{12}\\
h_{12}^{\dagger} & \mathcal{H}_{\mathrm{W}}^{*}\left(-\boldsymbol{k}\right)
\end{bmatrix},\label{eq:DirachamTB}
\end{eqnarray}
with
\begin{eqnarray}
h_{12} & = & -2c_{1}\sin k_{z}\left(\cos k_{x}-\cos k_{y}\right)\sigma_{x}\nonumber \\
 &  & +c_{2}\sin k_{x}\sin k_{y}\sin k_{z}\sigma_{y}.
\end{eqnarray}
This model gives a pair of Dirac points at $k_{z}$ axis and two Fermi
arcs on the (010) surface, which respectively occupying different
parts of the (010) surface BZ (see SM~\citep{Supplemental_Materials}). Thus, for the model (\ref{eq:DirachamTB}),
the shift current solely from both the bulk and the surface Fermi
arcs vanish. The calculated results of $\sigma_{zy}^{z}\left(\omega;y\right)$
for the (010) slab of the Dirac model (\ref{eq:DirachamTB}) are given
in Fig.~\ref{Fig.5}. Similar to the cases in WSM (\ref{eq:weylhamTB}),
$\sigma_{zy}^{z}\left(\omega;y\right)$ here also exhibits strong
surface enhancement and the distribution for $\sigma_{zy}^{z}\left(\omega;y=L/2\right)$
mainly concentrates around the two Fermi arcs, showing the surface
photocurrent in model (\ref{eq:DirachamTB}) also is induced by the
bulk-Fermi-arc transition.

The experimental detection of the surface shift current effect has
been well developed \citep{nc2019_Wang_exp_surfres,nc2017_huang_surface_spin_current},
and many materials are experimentally confirmed as topological (Weyl)
semimetals with surface Fermi arc \citep{
    rmp2021_Ding_Weyl,prl2017_Chang_optgeometry,
    nc2014_Cd3Ad2_expfermiarc,science2015_Na3Bi_expfermiarc,
    prl2015_TaAs_expfermiarc,np2016_Deng_MoTe2_expfermiarc,
    prl2019_CoSi_expfermiarc,prb2021_TaSeI_expfermiarc}.
Hence, the novel effect predicted here should be readily probed in experiments.

\begin{acknowledgments}
The authors thank S. A. Yang, Rui-Chun Xiao and J. Xun for helpful discussions. This work is supported by the National Key R\&D Program of China (Grants No.~2020YFA0308800), the NSF of China (Grants No.~11734003, No.~12061131002 and No.~12004035), the Strategic Priority Research Program of the Chinese Academy of Sciences (Grant No.~XDB30000000), and the Beijing Institute of Technology Research Fund Program for Young Scholars.
\end{acknowledgments}

\end{document}